\documentclass[showpacs,prb,twocolumn,floats,superscriptaddress]{revtex4}

\usepackage{graphicx}
\usepackage{amsmath}
\usepackage{amssymb}
\usepackage[colorlinks=true,citecolor=blue,linkcolor=blue]{hyperref}
\usepackage{amsfonts}
\usepackage[latin9]{inputenc}
\usepackage{color}
\usepackage{textcomp}

\renewcommand{\vec}[1]{\ensuremath{\boldsymbol{#1}}}

\begin{document}

\title{ Band gap engineering in  AA-stacked   bilayer graphene}
\date{\today }
\author{Hasan M. Abdullah}
\email{alshehab211@gmail.com}
\affiliation{Department of Physics, King Fahd University of Petroleum and Minerals, 31261 Dhahran, Saudi Arabia}
\affiliation{Saudi Center for Theoretical Physics, P.O. Box 32741, Jeddah 21438, Saudi Arabia}
\affiliation{Department of Physics, College of Applied Science, Taiz University, Yemen}

\author{Mohammed Al Ezzi}
\email{alezzi@u.nus.edu }
\affiliation{Department of Physics, National University of Singapore, 2 Science Drive 3, Singapore 117551, Singapore
}

\author{H. Bahlouli}
\affiliation{Department of Physics, King Fahd University of Petroleum and Minerals, 31261 Dhahran, Saudi Arabia}
\affiliation{Saudi Center for Theoretical Physics, P.O. Box 32741, Jeddah 21438, Saudi Arabia}

\begin{abstract}

We demonstrate   that   AA-stacked bilayer graphene (AA-BLG)  encapsulated by  dielectric materials can possess an energy gap   due to the induced mass term. Using the four-band continuum model, we evaluate transmission and reflection probabilities along with the respective conductances. Considering interlayer mass-term difference opens a gap in the energy spectrum and also couples the two Dirac cones. This cone coupling induces  an  inter-cone transport that is asymmetric with respect to the normal incidence in  the presence of asymmetric mass-term.
The energy spectrum of the  gapped AA-BLG exhibits    electron-hole asymmetry that is reflected in the associated intra- and inter-cone channels. We also find that  even though Klein tunneling exists  in gated  and biased AA-BLG,  it is precluded by the interlayer mass-term difference  and instead Febry-P\'erot resonances appear. 
\end{abstract}

\maketitle

\section{Introduction}
\textcolor{red}{ }The experimental realization of single graphene layers in 2004\cite{Geim_2007} have triggered tremendous interest in this material and its multilayer. The promising electrical, optical, mechanical properties as well as the great potential for sensor technology make graphene and its derivatives promising candidates for nanoscale device applications.  \cite{Castro_Neto_2009,Berdiyorov_2016,Tit_2017}. Graphene's extraordinary properties make it a promising candidate for  nanoscale device applications in the future. 
Moreover, bilayer graphene exists in two different types of stacking, namely, AB-(Bernal) and AA stackings (AB-BLG and AA-BLG). The stability of AB-BLG made it the subject of considerable investigations, both theoretical and experimental\cite{Ohta_2006,Goerbig_2011,Abdullah_2016}.  On the contrary to the previous belief that AA-BLG samples are unstable, recent stable samples were realized \cite{Lee_2008,Borysiuk_2011,de_Andres_2008,Liu_2009}.
The linear gapless energy spectrum of pristine AA-BLG attracted considerable theoretical interest \cite{Rakhmanov_2012,Mohammadi_2015,Chen_2013,Chiu_2013,Rozhkov_2011,Rozhkov_2016}. Among the existing studies on AA-BLG are spin Hall effect \cite{Dyrda__2014,Hsu_2010},    doping effects\cite{Sboychakov_2013}, dynamical conductivity\cite{Tabert2012}, tunneling through electrostatic and magnetic barriers\cite{Wang2013,Sanderson_2013}, magnon transport\cite{Owerre01_2016},influence of spin orbit coupling on the band structure\cite{Yao_2007}, and Landau levels in biased AA-BLG in the presence of nonuniform magnetic field\cite{Wang2013b}. 

The Klein tunnelling of Dirac fermions prevents the complete confinement in graphene. Overcoming this drawback can be achieved by opening a band gap in the energy spectrum by, for example, using slow Li$^+$ ions or perpendicular electric field in single-layer graphene and AB-BLG, respectively\cite{Oostinga_2007,Ohta_2006,Zhou_2007,Ryu_2016}. A different rout to achieve a perfect confinement in single layer graphene was through graphene quantum blisters\cite{Abdullah2018}where the charge carriers are confined on a delaminated bilayer graphene. 
In addition, substrates can also play a key role in the electronic confinement of single layer graphene due to the substrate-induced band gap of order of  $\backsim (20-500)$ meV\cite{Wang_2015,San_Jose_2014,Kindermann_2012,Song_2013,Jung_2015,Nevius_2015,Zarenia_2012}. The width of the band gap depends on the mass term induced by the substrate whose magnitude can be in the  order of $\backsim(50-100)$ meV depending on the type of the substrate \cite{Uchoa_2015}. It has been recently  showed that Hall phase can be realized in gapped AB-BLG when mass terms are considered in both layers\cite{Zhai2016}. Such mass terms are   induced  by dielectric materials such as hexagonal boron nitride (h-BN) or SiC. 

 A toy model suggested that a gap can be opened in AA-BLG if different spin-orbit coupling (SOC) are considered in each layer \cite{Prada_2011}. However, controlling the SOC in each layer is practicably not feasible  and most importantly the SOC in graphene is considered extremely small  which remains to be verified experimentally. For example, the  gap induced by SOC is  $0.8\times 10^{-5}$ meV for $\pi$ orbit and $9.0$ meV for the $\sigma$ orbit\cite{Yao_2007}.  Another study showed  that a band-gap in the transmission spectrum   of AA-BLG with double magnetic barriers which can be tuned by a bias\cite{Wang2013}.

In the present work we propose a configuration of AA-BLG that hosts a band gap in the energy spectrum arising from the   dielectric-induced mass terms. Considering the same mass terms in both layers of AA-BLG opens a gap around the lower and upper cones whereas the whole spectrum remains gapless. On the other hand, considering different mass terms in both layers breaks the inversion symmetry and, hence,  induces a gap in the energy spectrum. The width of the gap is directly affected by the inter-layer mass-term difference.
 Biasing the two layers of AA-BLG allows inter-cone transition due to the coupling of the upper and lower cones established by the bias. Such transition is forbidden in the case of zero bias but it can be also  induced when different mass terms are considered in both layers. 

This paper is organized as follows. In Sec. \ref{model}, we present the  model
of our study and describe the different transition processes  allowed in the system.
Sec. \ref{results} is devoted to numerical results and discussion  of conductance, transmission and reflection probabilities. Finally, we  conclude by stressing our main findings  in Sec. \ref{summary}. 
\section{electronic Model and energy spectrum}\label{model}
\begin{figure}[t]
\vspace{0.cm}
\centering\graphicspath{{./Figures/}}
\includegraphics[width=1.8  in]{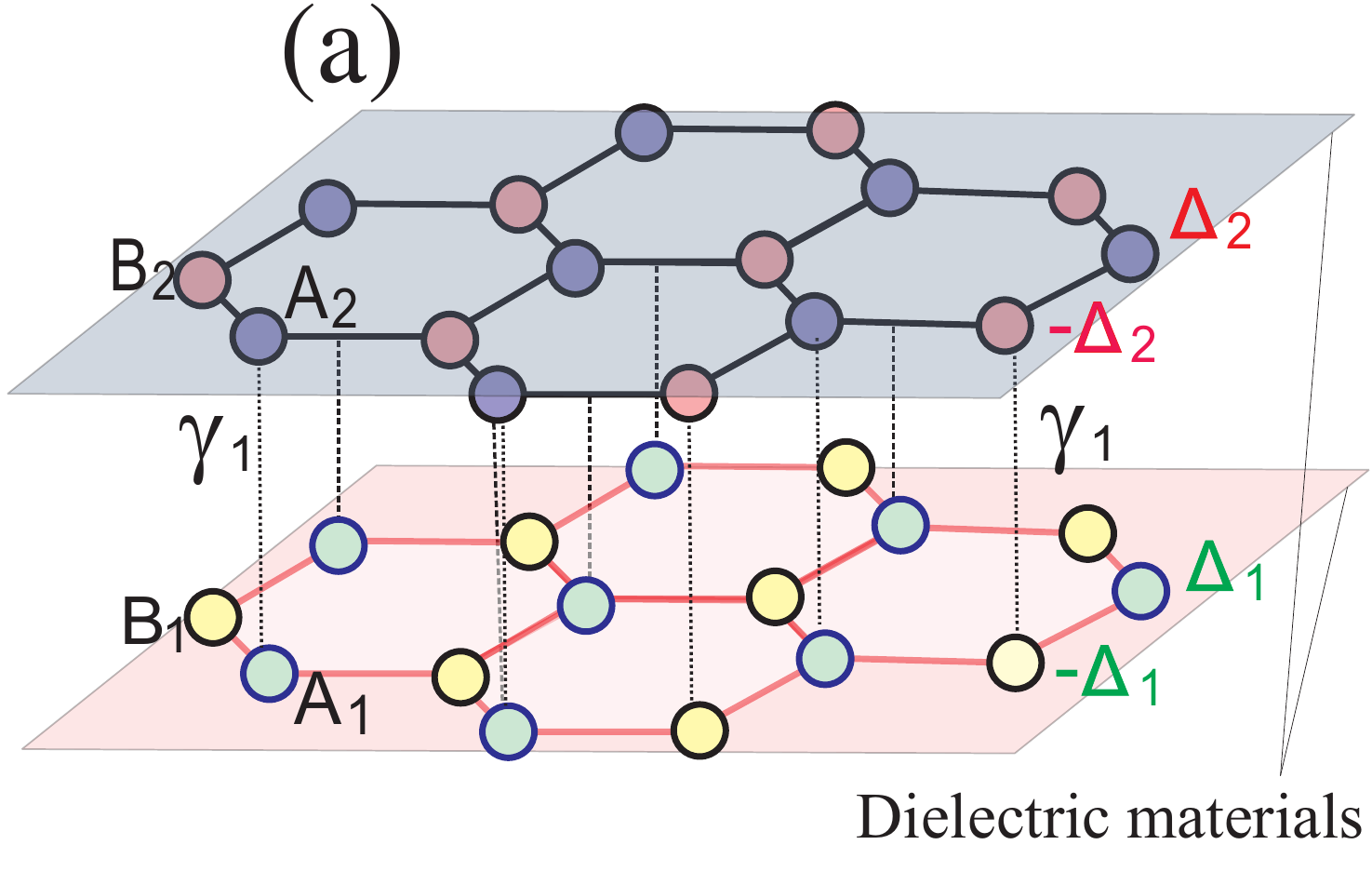}\ \ \
\includegraphics[width=1.4  in]{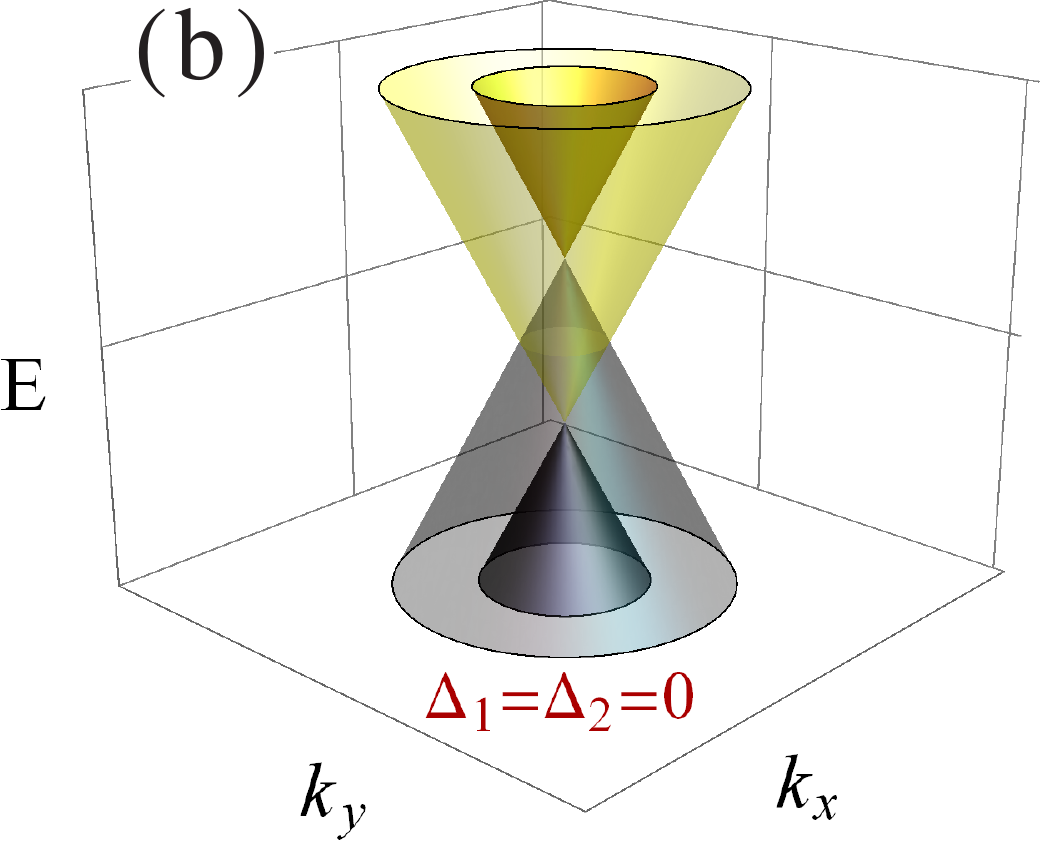}\\ \ \ \ \ \ 
\includegraphics[width=1.4  in]{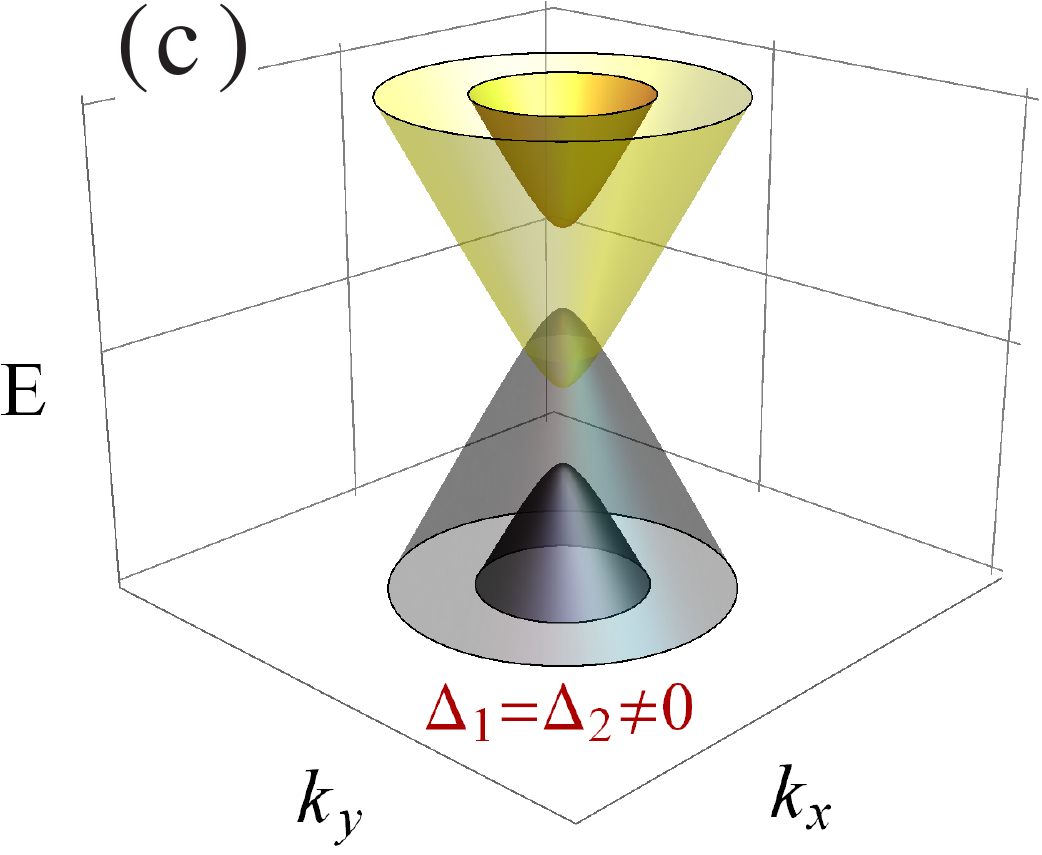}\ \ \ \ \ \ \
\includegraphics[width=1.4  in]{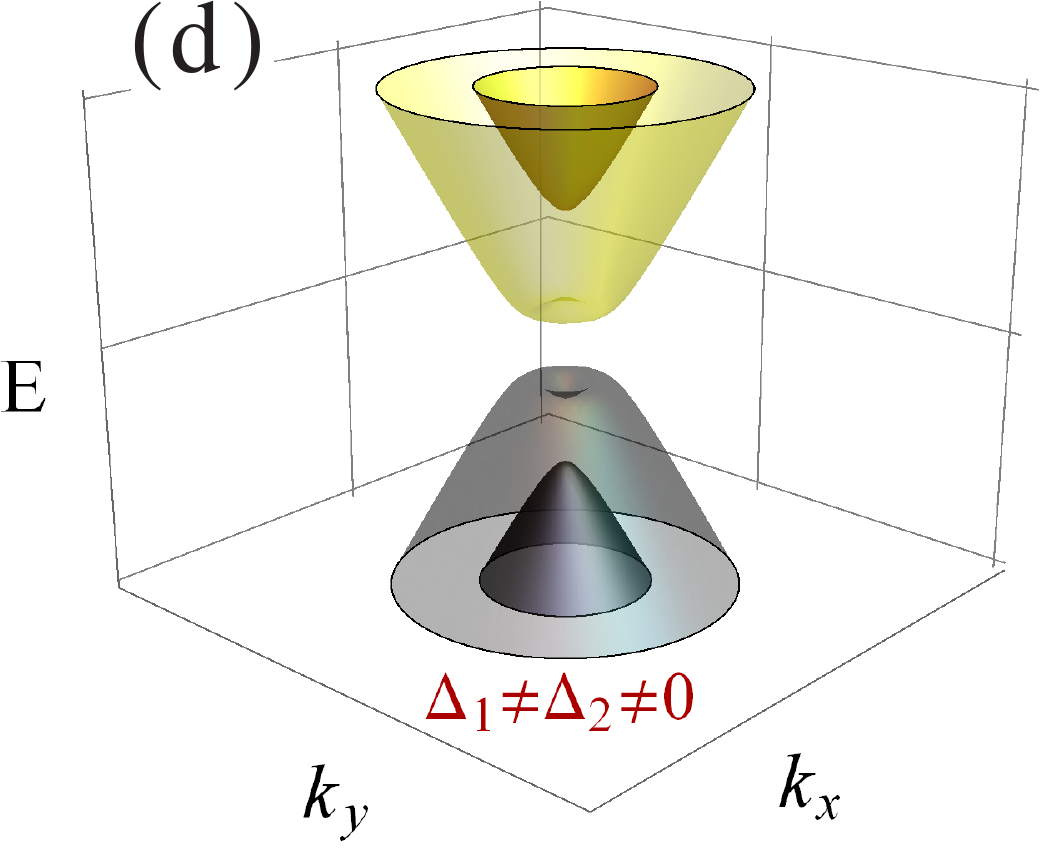}\
\vspace{0.cm}
\caption{(a) Crystalline structure of the AA-stacked  graphene bilayer  associated with the energy spectrum for: (b) zero mass-term amplitude, (c)  same mass-term in both layers,  and (d) with  different mass terms which induces a gap  in the spectrum.   }\label{Mass_Effect}
\end{figure}

Single layer graphene  has a hexagonal crystal structure  comprising two atoms $A$ and $B$ in its unit cell  whose interatomic distance $a=0.142$ nm and intera-layer coupling $\gamma_0=3$ eV\cite{Zhang_2011}. In the AA-stacked graphene bilayer the two single layer graphene are placed exactly on top of each other such that  atoms $A_2$ and $B_2$ in the top layer are located directly above the atoms $A_1$ and $B_1$ in the bottom layer,    with  direct inter-layer coupling $\gamma_1\approx0.2\ eV$ \cite{Lobato_2011}, see Fig. \ref{Mass_Effect}(a).  The energy spectrum of AA-BLG for different amplitudes of the mass terms  is shown in Figs. \ref{Mass_Effect}(b-d) where a gap  arises as a result of interlayer  mass-term difference. The presence of the gap in this case is a manifestation of  breaking the inversion  symmetry. In AA-BLG all atoms take part in the interlayer coupling contrary to the AB-BLG where only half of the atoms participate and  as a consequence  $\gamma_1^{AB}=2\gamma_1^{AA}\approx0.4$ eV\cite{Li2009,Xu_2010,Abdullah2017}. Another difference is that the latter  has asymmetric interlayer coupling, in other words, atom $A_1$  coupled to atom $B_2$  while the coupling is symmetric in the AA-BLG. Such differences give rise to the distinct  band structure and transport properties in both types of stackings.
The  continuum approximation of the Hamiltonian which describes the electrons near one of the Dirac points  $K$ or $K'$  of AA-BLG taking into account   SOC and mass terms   reads\cite{Tabert2012} 
\begin{equation}\label{eq01}
H^{\tau}=\left[
\begin{array}{cccc}
  H^{\tau}_{1}& \gamma_{1} I \\
  \gamma_{1} I  &  H^{\tau}_{2}
\end{array}%
\right],
\end{equation}
where $ H_{i}^{\tau}=\tau  v_{F}\left( p_{x}\sigma_{x}+\tau p_{y} \sigma_{y} \right)+\tau\sigma_{z}(s_{z}\lambda_{i}+\Delta_{i}) +V_{i} I $ is the single layer graphene  Hamiltonian with $V_{i}$  the electrostatic potential, whose width is $d$, on the $i$-th layer which can be varied using top and back gates on the sample and $v_{F}=10^{6}$ m/s the charge carries  speed in the graphene sheet, $p_{x,y}=-i\hbar \partial_{x,y}$ ,  $\sigma_{x,y}$ and  $I$ are the $2\times2$ Pauli and identity matrices, respectively. The strength of the intrinsic SOC and the mass term,  in the  $i$-th layer are represented by $\lambda_i$ and $ \Delta_i$   , respectively,  $\tau=\pm1$ corresponds to the $K$ and $K'$ valleys and $s_{z}=\pm1$ stands for the electron spin up and spin down. When considering only spin up $s_{z}=1$  it becomes clear  that the mass term plays exactly the same role as the SOC in the single layer Hamiltonian. Being extremely small, SOC has insignificant effect on the band structure and transport properties specially at high energies therefore  it will be neglected in the  further calculations  of transmission, reflection and conductance. A simplification can be made to the Hamiltonian, in the vicinity of $K$ valley, by applying unitary transformation that forms symmetric and anti-symmetric combination of the top and bottom layers. This results in a Hamiltonian in the basis $\mathbf{\Psi}=2^{-1/2}(\Psi_{A 2}+\Psi_{A 1},\Psi_{B 2}+\Psi_{B 1},\Psi_{A 2}-\Psi_{A 1},\Psi_{B 2}-\Psi_{B 1})^{T}$   of the form:
\begin{widetext}
\begin{equation}\label{eq02}
H=\left(
\begin{array}{cccc}
  \gamma_1+v_0+\Delta_0 & v_{F}\pi^{\dag} & \delta+\Omega & 0 \\
  v_{F}\pi &  \gamma_1+v_0-\Delta_0 &  0 & \delta-\Omega\\
  \delta+\Omega &   0 &  -\gamma_1+v_0+\Delta_0& v_{F}\pi^{\dag} \\
  0 & \delta-\Omega& v_{F}\pi & -\gamma_1+v_0-\Delta_0 \\
\end{array}%
\right).
\end{equation} 
\end{widetext}    
where$\ v_0=(V_2+V_1)/2,\ \delta=(V_2-V_1)/2,\ \Delta_0=(\Delta_2+\Delta_1)/2$, and $ \Omega=(\Delta_2-\Delta_1)/2$.
Introducing the length
scale $l=\hbar v_{F}/\gamma_{1}$, which represents the inter-layer coupling length $l\approx3.3$ nm, allows
us to define the following dimensionless quantities:
\begin{widetext}
\begin{align*}
E\rightarrow\frac{E}{\gamma_1},\ v_0\rightarrow\frac{v_0}{\gamma_1},\ \delta\rightarrow\frac{\delta}{\gamma_1},\ \Delta_0\rightarrow\frac{\Delta_0}{\gamma_1},\ \Omega\rightarrow\frac{\Omega}{\gamma_1},\  k_y\rightarrow lk_y,\ \text{and}\ \vec r\rightarrow\frac{\vec r}{l}.
\end{align*}
\end{widetext} 
 As a result of the  translational invariance along the $y$ direction,  the momentum    in that direction is  a conserved quantity and, hence, the wavefunction in the new basis can be written as
\begin{equation}\label{eq03}
\mathbf{\Psi}(x,y)=e^{iyk_y}\left[\phi_{1},\phi_{2},\phi_{3},\phi_{4}\right]^{\dag},
\end{equation} where $\dag$ stands for the transpose.  Implementing Schrodinger equation  $H\Psi=E\Psi$   leads to  four coupled differential equations:
\begin{eqnarray}
&&-i\left[\frac{d}{dx}+ky\right]\phi_{2}+(\delta+\Omega)\phi_{3}=(\epsilon-1-\Delta_0)\phi_{1}\label{eq04}\\
&&
-i\left[\frac{d}{dx}-ky\right]\phi_{1}+(\delta-\Omega)\phi_{4}=(\epsilon-1+\Delta_0)\phi_{2}\label{eq05}\\
 &&-i\left[\frac{d}{dx}+ky\right]\phi_{4}+(\delta+\Omega)\phi_{1}=(\epsilon+1-\Delta_0)\phi_{3}\label{eq06}\\
&&-i\left[\frac{d}{dx}-ky\right]\phi_{3}+(\delta-\Omega)\phi_{2}=(\epsilon+1+\Delta_0)\phi_{4}\label{eq07}
\end{eqnarray}
   where $\epsilon=E-v_0$. 
The  system of coupled first-order differential equations (\ref{eq04}-\ref{eq07})  can be transformed into a single second order differential equation  for $\phi_{2}$ as follows  
\begin{equation}\label{eq08}
\left[\frac{d^{2}}{dx^{2}}+(k^{\pm}_x)^{2}\right]\phi_{2}=0,
\end{equation}
where
\begin{equation}\label{eq09}
k_x^{\pm}=\left[-k^{2}_{y}+\beta^{2}
\pm \sqrt{\xi^2+4(\epsilon^2-\Omega^{2})}\right]^{1/2},
\end{equation}
with $\beta^{2}=1+\epsilon^2+\delta^2-\Delta_0^2-\Omega^2$ and $\xi=2(\epsilon \delta+\Delta_0 \Omega)$. From Eq. \eqref{eq09}, it follows that the energy spectrum for the system is given by
\begin{equation}\label{eq10}
\epsilon_\alpha^{\pm}=\frac{1}{2}\left[\alpha\sigma
\pm \sqrt{-(\sigma^2+2\rho)-2 \textrm{sgn}(\alpha)\frac{\kappa}{\sigma}}\right]^{1/2},
\end{equation}
where $\alpha=\pm1$ and $\sigma,\ \rho$ , and $\kappa$  functions are defined in   Appendix \ref{Appe-A}. In  Fig. \ref{Bands}. we show the energy spectra of the AA-BLG for different values of the system parameters. For $\delta=0$, the  characterized quantities $\Delta E,\ \Delta k_y,\ \Delta E_m$, and  $E_g$  in Fig. \ref{Bands} are defined as follows
\begin{subequations}
\begin{eqnarray}
&&\Delta E=2\left\vert \sqrt{1+\Omega^2}-\Delta_{0} \right\vert,\label{eq11-a}\\
&& \Delta k_y=2\sqrt{1-\Delta_0^2\left( 1+\Omega^2 \right)},\label{eq11-b}\\
 &&\Delta E_m=2\Delta_0,\label{eq11-c}\\
 &&E_g=2\sqrt{\left( 1-\Delta_0^2 \right)}\left\vert \Omega \right\vert.\label{eq11-d}
\end{eqnarray}  
 \end{subequations}
 
AA-BLG has a linear energy  spectrum  with two up-down  Dirac cones   shifted by $\Delta E$, which is $2\gamma_1$ in this case as shown in Fig. \ref{Bands}(a) by the solid black curves. When AA-BLG is subjected  to a perpendicular electric field (biased AA-BLG) the two Dirac cones are   slightly shifted and  situated at $v_0\pm\sqrt{\gamma_1^2+\delta^2}$ , see dotted-dashed orange curves in Fig. \ref{Bands}(a). Fig. \ref{Bands}(b) is the same as Fig. \ref{Bands}(a) but with the same mass term amplitude $\Delta_0=0.5\gamma_1$ for both layers and the spectrum exhibits a shift $\Delta E_m$ in the bands   around the upper and lower Dirac cones.
\begin{figure}[t]
\vspace{0.cm}
\centering\graphicspath{{./Figures/}}
\includegraphics[width=3.5  in]{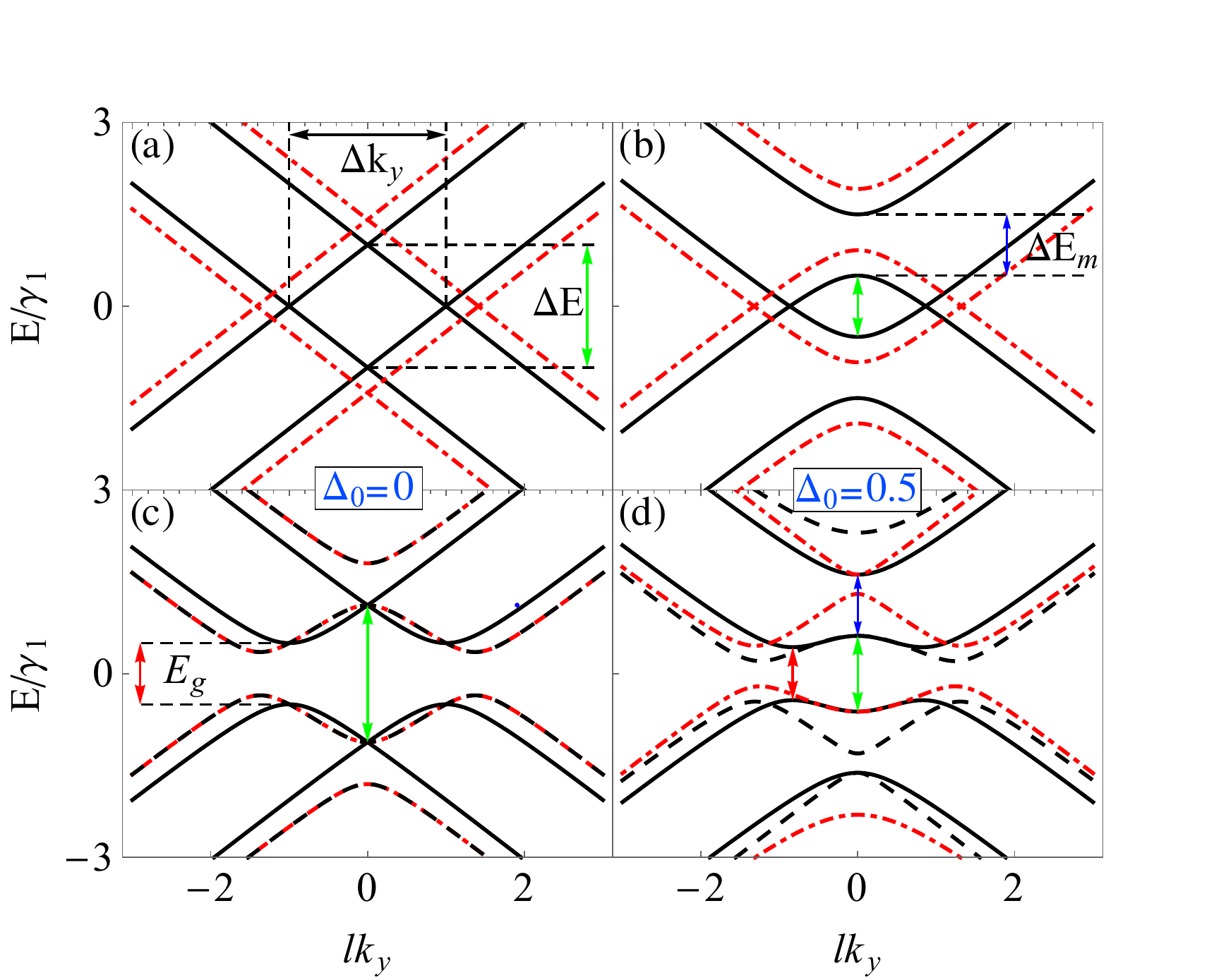}
\caption{Energy spectrum of AA-stacked bilayer graphene with (a)  no mass term been considered, (b)  the same mass terms in both layers, (c) different mass terms in top and bottom layers, and (d) same magnitude and different sign in each layer (solid black curves). The dashed orange and black curves correspond to the same system but  with  electric fields in opposite  directions  with $\delta=(1,-1)\gamma_1$, respectively. }\label{Bands}
\end{figure}
 Introducing a mass term difference $\Omega=0.5\gamma_1$ with mass term amplitude $\Delta_0=0.5\gamma_1$  leads to opening  a gap $E_g$  in the energy spectrum as shown in  Fig. \ref{Bands}(c)  where solid black, dotted-dashed orange, and dashed black curves correspond to $\delta=(0,1,-1)\gamma_1$, respectively.
Notice that considering biased AA-BLG ($\left\vert \delta \right\vert\neq0$) with inter-layer mass-term difference and non-zero amplitude, i.e. $(\Omega,\Delta_0)\neq0$,  breaks the electron-hole symmetry as indicated by the black dashed  and dotted-dashed orange curves in Fig. \ref{Bands}(d). The energy spectrum exhibits another symmetry which can
be obtained under the exchange $[E(k_y),-\delta]\leftrightarrow[-E(k_y),\delta]$. In contrast, keeping the mass-term difference $\Omega=0.5\gamma_1$ but with zero amplitude $\Delta_0=0$ restores the electron-hole symmetry under the exchange $\delta\leftrightarrow-\delta$ without affecting the existence   of the   energy gap $E_g$ , see Fig. \ref{Bands}(c).

 \begin{figure}[t]
\vspace{0.cm}
\centering\graphicspath{{./Figures/}}
\includegraphics[width=3.  in]{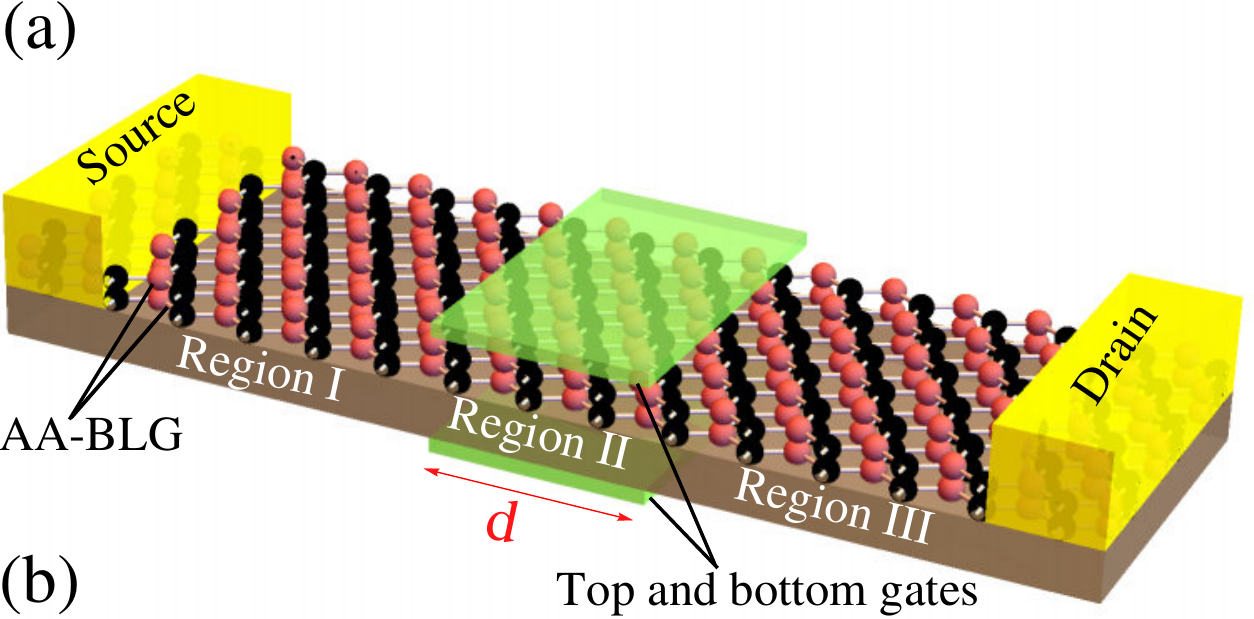}\\  
\includegraphics[width=3.  in]{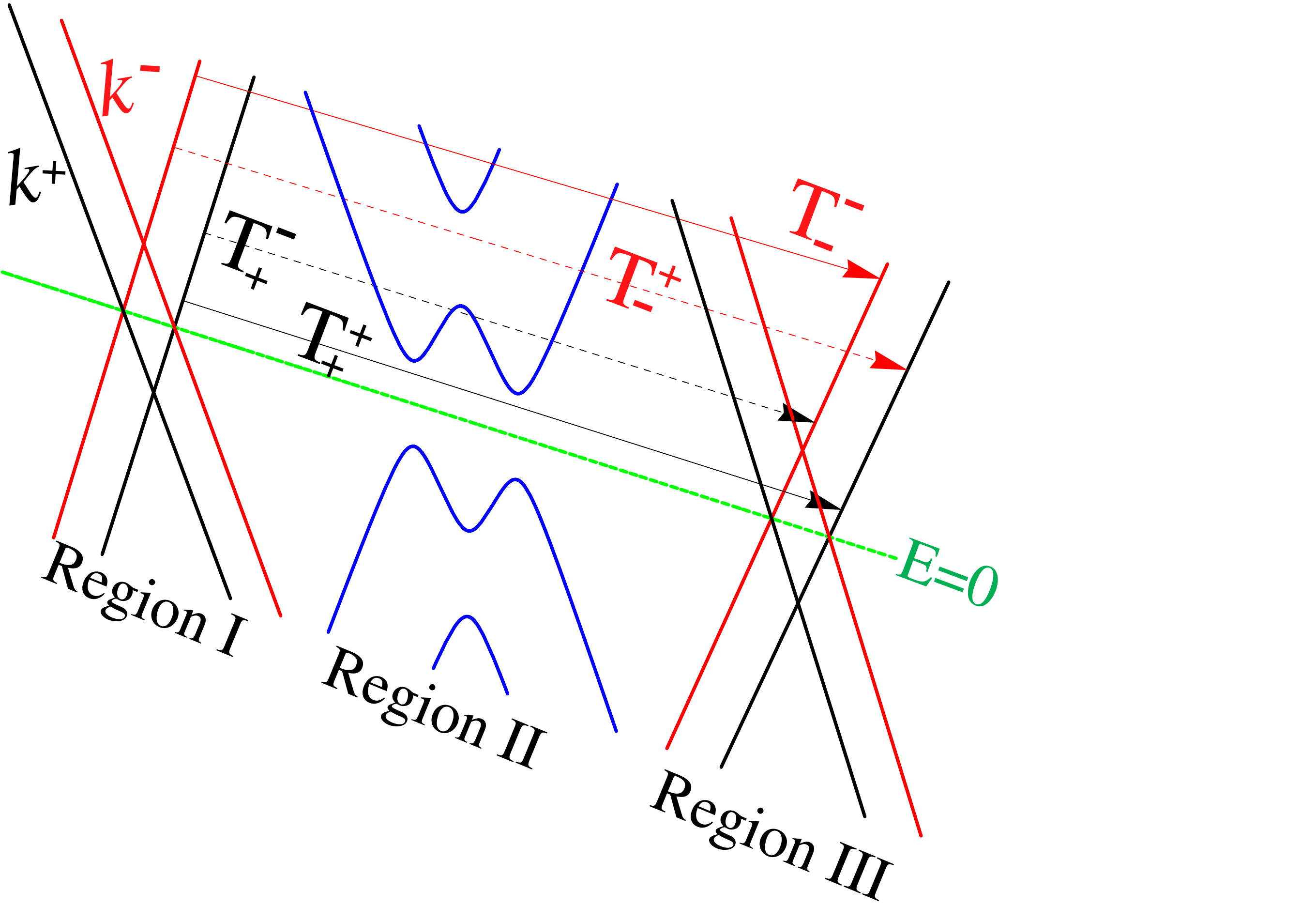}\\ 
\vspace{0.7cm}
\includegraphics[width=1.6  in]{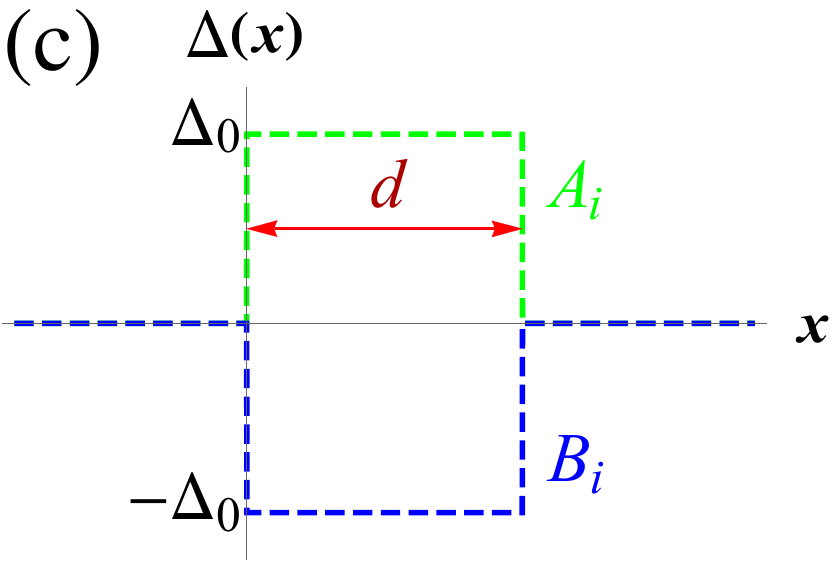}\
\includegraphics[width=1.6  in]{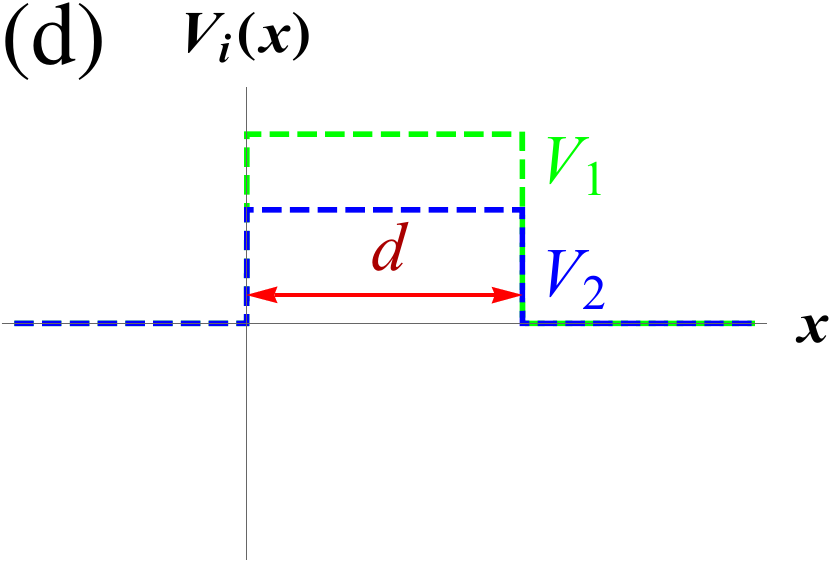}\
\vspace{0.cm}
\caption{(a) Schematic representation of the proposed system with the parameters of the electrostatic  rectangular potential on each layer of AA-BLG. (b) The mass term profile on  sublattices $A_i$ and $B_i$ in the $i$-th layer. (c) Electrostatic potential strength $V_i$ applied to the $i$-th layer. }\label{Device}
\end{figure}
\begin{figure}[t]
\vspace{0.cm}
\centering\graphicspath{{./Figures/}}
\includegraphics[width=3.35  in]{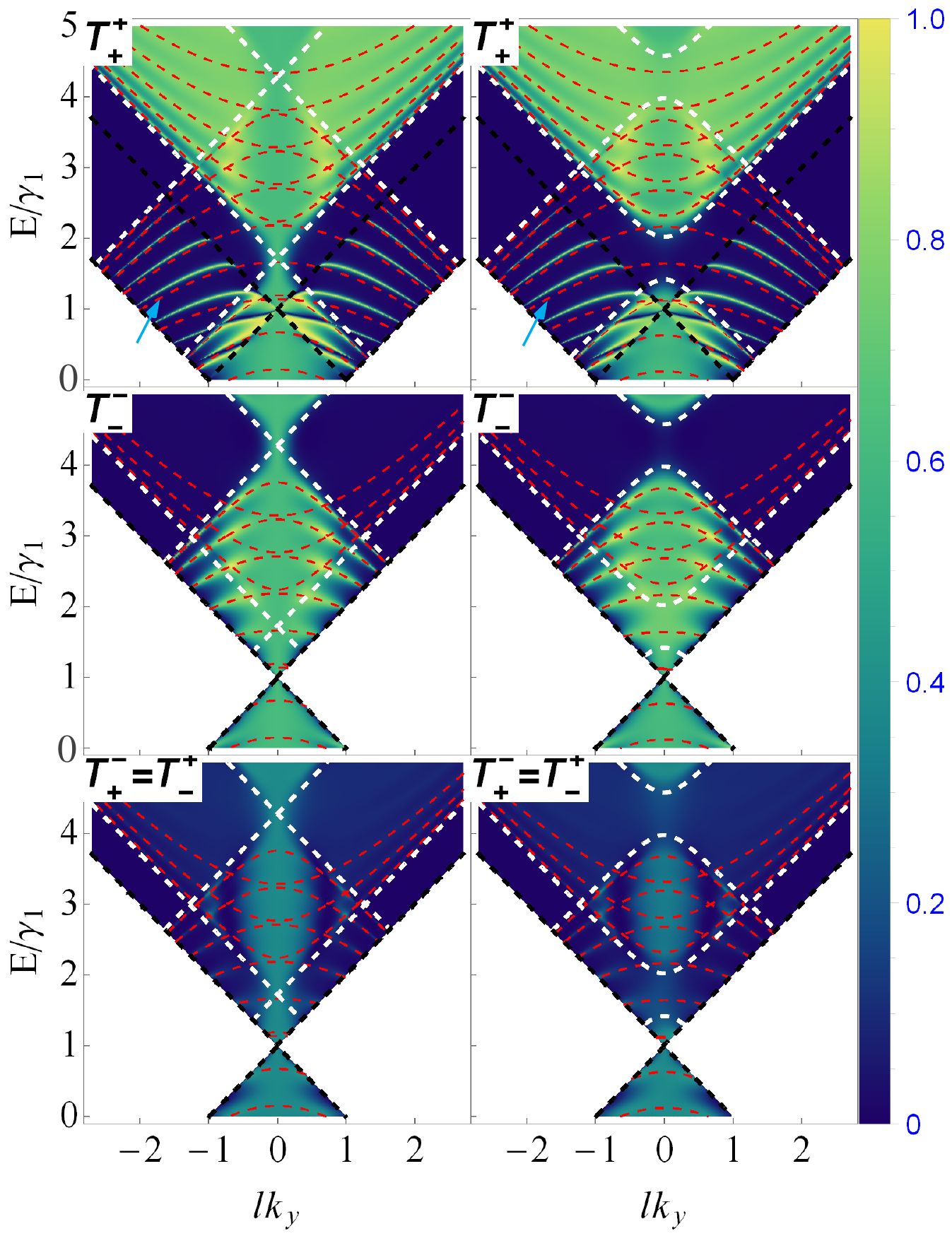}
\vspace{0.cm}
\caption{Density plot of the transmission probabilities for   $ \Omega=0,\ v_0=3\gamma_1,\ d=6l,\ \delta=0.8\gamma_1$, with $\Delta_0=0$ (left panel) and  $\Delta_0=0.3\gamma_1$(right panel). The superimposed  dashed black and white curves correspond to the bands outside and inside the barrier regions, respectively. The red dashed curves correspond to the Febry-P\'erot resonances given by Eq. \eqref{resonances}.}\label{T_Zero_Om}
\end{figure}
\begin{figure*}[t!]
\vspace{0.cm}
\centering\graphicspath{{./Figures/}}
\includegraphics[width=6  in]{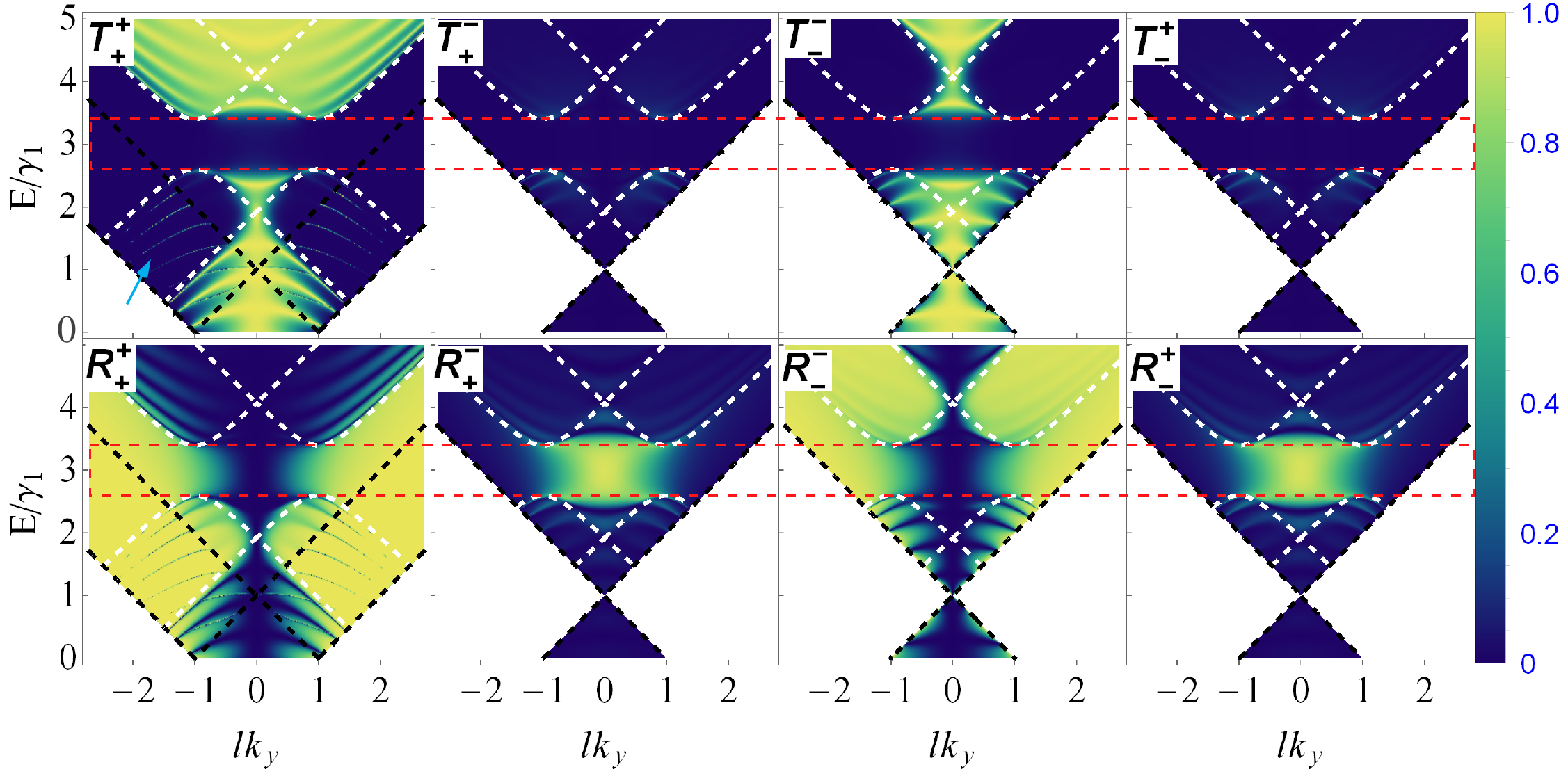}
\vspace{0.cm}
\caption{Density plot of the transmission and reflection probabilities for   $ \Omega=0.4\gamma_1,\ v_0=3\gamma_1,\ d=6l,\ \delta=0$, with $\Delta_0=0$. The superimposed  dashed black and white curves correspond to the bands outside and inside the barrier regions, respectively.    }\label{T_R_Symmetric}
\end{figure*}

\subsection{Transitions probabilities}
 To investigate the quantum transport of the proposed system in Fig. \ref{Device}(a), we first need to find the wavefunction in each region , see Appendix \ref{Appe-B} for more details. In the first $I$ $(x<0)$ and third $III$ $(x>d)$ regions we have pristine AA-BLG. While in the intermediate region $II$ $(0<x<d)$ the mass term induced by the dielectric materials as well as electrostatic potential is included. The induced mass-term on sublattices $A_i$ and $B_i$  has the same amplitude but different sign as shown in  Fig. \ref{Device}(c), while the electrostatic potentials applied to top and bottom layers have the same effect on both sublattices, see   Fig. \ref{Device}(d).
 In AA-BLG, there are two propagating modes for all energies in contrast to AB-BLG whose two modes only exist for energies exceeding the interlayer coupling. These two modes $k^+$ and $k^-$exist outside the interaction region $III$ and correspond to the lower and upper Dirac cones, respectively, see Fig. \ref{Device}(b).   Hence, we need to consider the different transition probabilities between the lower and upper cones. There are four different transmission probabilities  as illustrated  in  Fig. \ref{Device}(b). We label a transition from one cone to another by $A_i^j$ which denotes a charge carrier scattering from the cone $k^i$ to the cone $k^j$, where $A$  can stand for transmission ($T$) or reflection ($R$) probabilities with $i,j=\pm$.  In  Appendix \ref{Appe-B} we explain how to obtain these probabilities. The  diagonal blocks of the Hamiltonian
in Eq.\eqref{eq03} represents two cones shifted in energy by $2\gamma_1$ (for $\Delta_0=0$) while the off-diagonal blocks characterize the coupling strength between the two cones. Therefore, for $\Omega=\delta=0$ the Hamiltonian stands for two decoupled cones where scattering between them is strictly  forbidden.
The scattering between these two cones can be induced only by the electrostatic potential bias $\delta$ or the  interlayer mass-term difference $\Omega$. The different processes of the inter- or intra-transition between the lower and upper cones are shown in Fig. \ref{Device}(b). Consequently, considering the scattering process between the two cones results in four channels for the transmission $T_\pm^j$  as well as for the reflection $R_\pm^j$ probabilities. For normalization  consideration we have     
\begin{equation}  
  \sum_{j=\pm}\left( T_\pm^j+R_\pm^j \right)=1.
\end{equation}  
For example, for the lower cone we have $T_+^++R_+^++T_+^-+R_+^-=1$. The zero temperature conductance can be calculated using the B\"uttiker's formula\cite{Snyman_2007,Blanter_2000}  \begin{equation}  
 {G_i^{j}}(E)=G_{0}\frac{L_y}{2
\pi}\int_{-\infty}^{+\infty}dk_{y} T_{i}^j(E,k_y),
\end{equation}
with $(i,j)=\pm$,  $L_y$  the length of the sample in the $y$-direction, and
$G_0=4\ e^2/h$. The factor $4$ comes from the valley and
spin degeneracy in graphene.
 The total conductance of the system  is the sum through all available channels $G_T=\sum_{i,j}G_i^j$. 

\section{Numerical results and discussion}\label{results}

As discussed in the preceding sections, scattering between the lower and upper cones can be  induced only by inter-layer bias or inter-layer mass difference. The effect of  the mass-term amplitude on the transport properties and Klein tunneling in unbiased\cite{Redouani2016} and biased\cite{Abdullah2018a} AA-BLG has been studied before.  Here we briefly discuss the results of a biased AA-BLG in the presence of the mass-term amplitude  to compare  the strength of inter-cone scattering  with the one arisen by the interlayer mass-term difference. In Fig. \ref{T_Zero_Om} we show the density plot of the different transmission channels as a function of the Fermi energy and the wave vector  $k_y$ for $\Delta_0=0\ (0.3)\gamma_1$  in left (right) panel.
We notice that the presence of the mass-term amplitude completely suppresses the intra-cone transmission ($T_+^+$ and $T_-^-$) in the vicinity of the upper and lower cones ($v_0-(+)\gamma_1+\Delta_0\geqslant E\geqslant v_0-(+)\gamma_1-\Delta_0$) as depicted in Fig. \ref{T_Zero_Om} (right panel). On the other hand,  the profile of the inter-cone transmission  $(T_-^+=T_+^-)$ shows finite transport within the same energy ranges rendering charge carriers  in the system  unconfined\cite{Abdullah2018a}.
 The propagating modes interfere with themselves in the interaction region as a result of the finite size effect. This interference leads to oscillation in the transmission probabilities at quantized energies. For $\Omega=0$ this so-called Febry-P\'erot resonances\cite{Snyman_2007}   appear at
\begin{widetext}  
\begin{equation}\label{resonances}
E_{\alpha,n}^\pm(k_y)=v_0+\alpha\sqrt{1+k_y^2+\delta^2+\Delta_0^2+\left( \frac{n\pi}{d} \right)^2\pm\sqrt{\left(1+\delta^2\right)\left( k_y^2+\Delta_0^2+\left( \frac{n\pi}{d} \right)^2 \right)}},
\end{equation}
\end{widetext}

where $\alpha=\pm1$. These energies are superimposed as red dashed curves in Fig. \ref{T_Zero_Om}. The resonances given by Eq. \ref{resonances} are valid only when the modes inside and outside the barrier are propagating as can be inferred from  the intra- and inter-cone channels $T^+_+$ and $T^-_-$. For example, the resonances marked by the blue arrow in the  channel $T^+_+$  in Fig. \ref{T_Zero_Om} do not obey Eq. \ref{resonances}. Such resonances occur in regions where the $k^+$ mode inside the barrier is evanescent while it is propagating outside.   They are arising  as a result of the cones coupling established by the bias, see Eq. \ref{eq02},  and they completely vanish once the two cone are decoupled\cite{Abdullah2018a}.

So far, we have assumed that the interlayer mass-term difference is zero in the interaction region.
To thoroughly examine its effect on the intra- and inter-cone transport, we show in Fig. \ref{T_R_Symmetric} the corresponding transmission and reflection probabilities in the presence of a symmetric mass-term amplitude, i.e.  $\Delta_0=0$ while $\Omega=0.4\gamma_1$. The different probabilities are plotted as a function of the conserved wave vector $k_y$ and the Fermi energy $E$.
The white zone in Fig. \ref{T_R_Symmetric} indicates the absence of the relevent propagating mode in the incident region which coincides with evanescent waves.
At first glance, we note that all intra- and inter-cone channels in transmission and reflection  are invariant under $k_y\rightarrow-k_y$ rendering it  symmetric with respect to normal incidence. This is a manifestation of the symmetric mass term introduced on  both layers as well as the symmetric inter-layer coupling in AA-BLG.  These symmetries also result in a symmetric inter-cone transport such that $T_+^-=T_-^+$ and $R_+^-=R_-^+$ as can be inferred from Fig. \ref{T_R_Symmetric}. Due to the absence of the propagating modes in the interaction region that are indispensable for tunneling, the intra- and inter-cone      transmission have been completely suppressed within  the induced gap, i.e. in the range $v_{0}+E_{g}/2\geqslant E\geqslant v_0- E_g$/2, with $E_g=0.8 \gamma_1$ as expressed in Eq. \eqref{eq11-d}. Furthermore, we notice also that the major transmitted current is carried out through the intra-cone  channel $T_+^+$. This implies that the cone coupling established by mass-term difference  is weak compared to the one induced by the potential bias. The coupling effect can be also   seen in the fringes within the domain marked by the blue arrow in the $T^+_+$ channel. They are attenuated  and sharper than  those   in Fig. \ref{T_Zero_Om}. On the other hand, we find that the reflection  profiles  display a slightly different   features as clarified in Fig. \ref{T_R_Symmetric}. Of particular significance is the remarkable observation of the extremely strong  inter-cone reflection $R_i^j$  within the energy  gap, specially near the normal incidence direction. In addition, for normal incidence the inter-cone reflection is finite, contrary  to the inter-cone transmission which is strictly zero. This is
a consequence of the fact that the  inversion and chiral symmetries are broken in the presence of the interlayer mass-term difference.

\begin{figure}[t!]
\vspace{0.cm}
\centering\graphicspath{{./Figures/}}
\includegraphics[width=\linewidth]{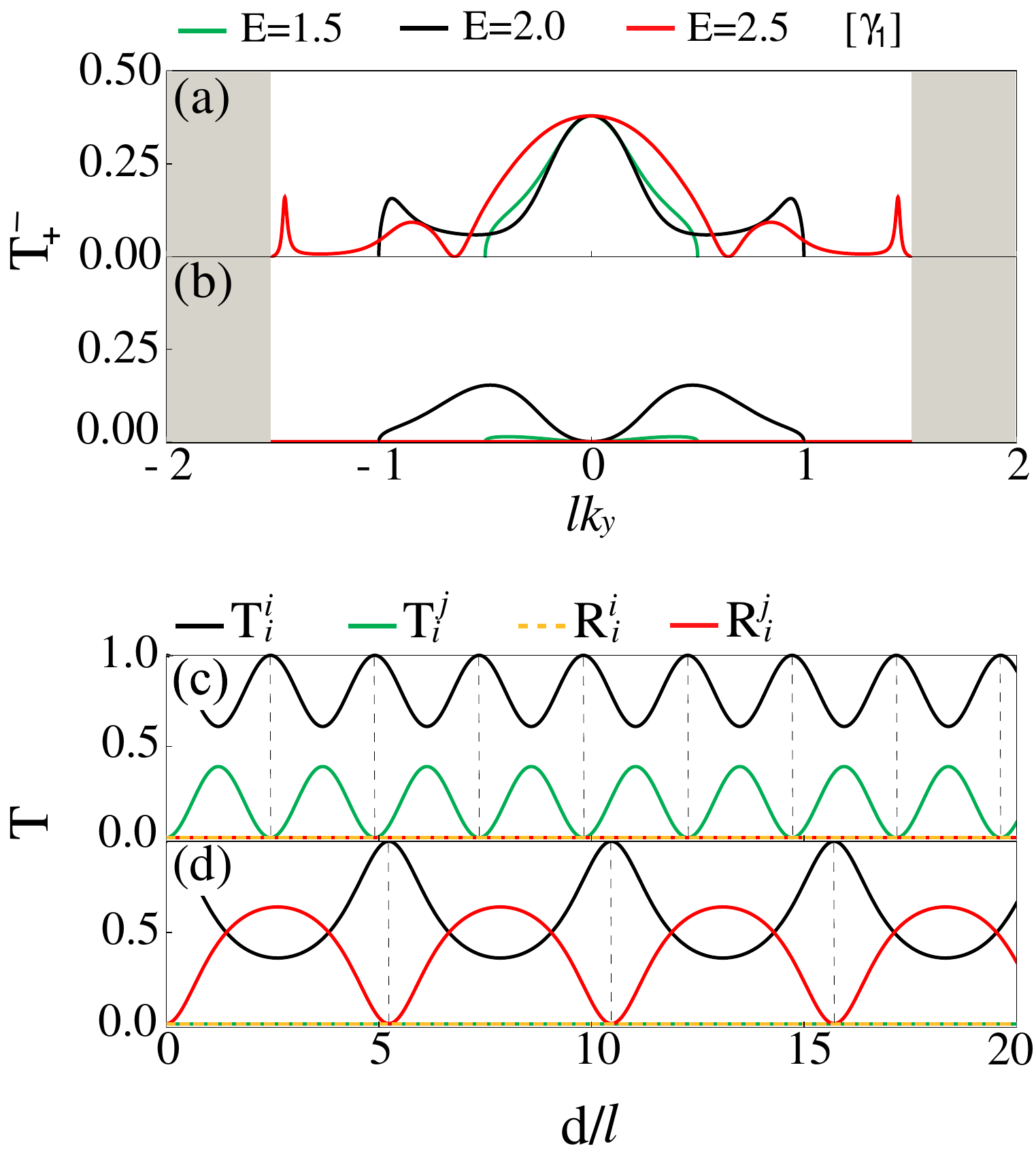}\\
\vspace{0.cm}
\caption{(a, b) Inter-cone transmission as a function  of the transverse wave vector $k_y$  at different energies with $d=6l$, $v_{0}=3\gamma_{1}$, $\Delta_{0}=0$. We consider only potential bias in panel (a), i.e. $\delta=0.8\gamma_1$,  $\Omega=0$, and interlayer mass-term difference in panel (b), i.e. $\delta=0$,  $\Omega=0.8\gamma_1$. The gray zones in panel (a) indicate the absence of the
relevent propagating mode in the incident region.  (c, d) Intra- and inter-cone channels for normal incidence as a function of interaction region width $d$, the parameters are set the same as in panels (a, b), respectively.  }\label{T_vs_d}
\end{figure}

As an illustration of the mass-term  influence on chiral tunnelling properties and the cone coupling, we show the transmission as a function of the wave vector $k_y$ and the width of the interaction region $d$ in Figs. \ref{T_vs_d}(b, d), respectively.  It is also  instructive to compare it with the case of biased AA-BLG to elucidate the effect of bias   and to give a deeper understanding of the cone coupling,   as depicted in Fig. \ref{T_vs_d}(a, c). First we consider the cone coupling established by the potential and mass-term difference and show the   inter-cone transmission  in Figs. \ref{T_vs_d}(a, b), respectively, for different Fermi energies.   Fig. \ref{T_vs_d}(a) shows the inter-cone transmission induced as a result of the cone coupling established by the interlayer potential bias.
We note  that with increasing the Fermi energy of the incident charge carriers, the strength of the associated maxima is invariable. Thus, it is clear that this maximum does not depend on the incident Fermi energy but rather on the bias strength and the width of the interaction region as will be clarified latter. On the other hand, from Fig. \ref{T_vs_d}(b) one can see that for normal incidence the inter-con transmission induced by the interlayer mass-term difference is strictly zero.
It is significant around the edges of the gap and vanishes at low or high energies. The fact that at normal incidence the inter-cone transmission is finite and zero with interlayer bias and mass-term difference, respectively, can be attributed to the chiral symmetry in the system. Introducing a bias dose not break the chiral symmetry in the system, but it is broken in the case of the interlayer mass-term difference. This suggests that the Klein tunneling should hold in the presence of the electrostatic bias and disappear with introducing the interlayer mass difference. 
\begin{figure*}[tb]
\centering\graphicspath{{./Figures/}}
\includegraphics[width=6  in]{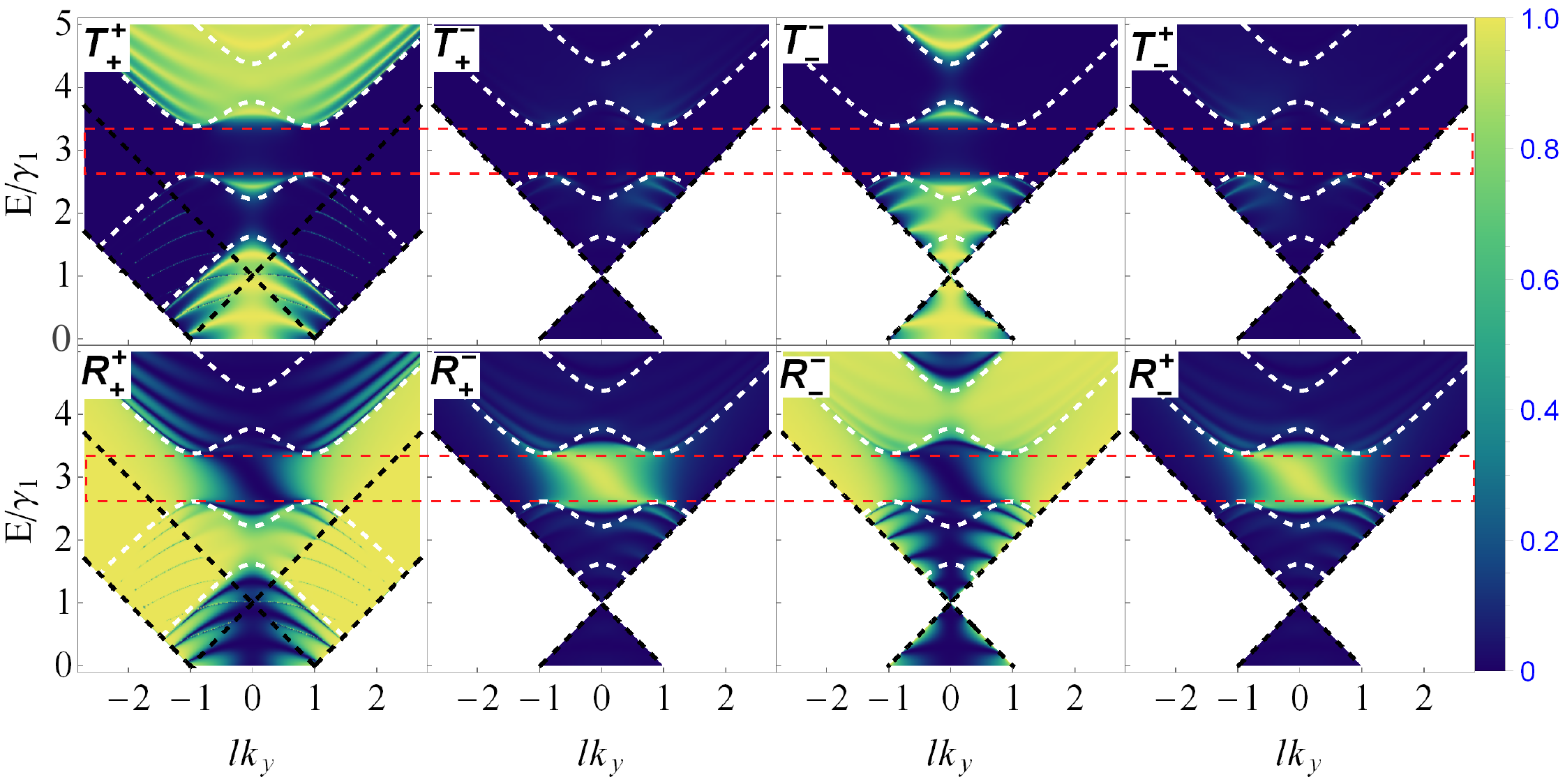}\
\caption{Density plot of the transmission and reflection probabilities for   $ \Omega=0.4\gamma_1,\ v_0=3\gamma_1,\ d=6l,\ \delta=0$, with  $\Delta_0=0.3\gamma_1$. The superimposed  dashed black and white curves respectively correspond  to the bands outside and inside the barrier regions.}\label{T_R_Asymmetric}
\end{figure*}

To elucidate how the Klein tunneling would be affected by the  bias and the interlayer mass-term difference, we show in Figs. \ref{T_vs_d}(c, d) the intra- and inter-cone transmission and reflection for normal incidence as a function of the interaction region width $d$. Before proceeding, we would like to remind the readers that we have two channels corresponding to the lower and upper cones. Each channel is normalized to unity and thus to observe Klein tunneling the total transmission and refection of each channel must by unity and zero, respectively. Fig. \ref{T_vs_d}(c) shows  all channels in case of finite bias and zero mass-term difference. It reveals that the intra- $T_i^i$ and inter-cone $T_i^j$ transmissions oscillate in anti-phase with  increasing the width of the interaction region, while both intra- and inter-cone reflections are zero. The maxima associated with intra- and inter-cone transmission coincide with the width
\begin{equation}
d_n=\frac{\pi(n+\eta)}{\sqrt{1+\delta^2}}
\end{equation}
with $n=0,1,2..$., and $\eta=(0,1/2)$ for  $T_i^i$ and $T_i^j$, respectively.  Such oscillation was also observed within domain walls in   delaminated  bilayer graphene\cite{Abdullah_2016}. Note that the location of these maxima is independent of the Fermi energy. Moreover, we can clearly now see that $T_i^i+T_i^j=1$ and $R_i^i+R_i^j=0$ regardless of the width of the interaction region $d$. This is a quintessential  trait of Klein tunneling in the system. Consequently, the system retains  Klein tunneling even in the presence of the potential bias. On the contrary,
in the presence of the interlayer mass-term difference
the intra-cone transmission $T_i^i$ and inter-cone reflection
$R_i^j$ are finite while $T_i^j$ and $R_i^i$ vanish, see Fig. \ref{T_vs_d}(d). Again, we see also here that the none zero channels $T_i^i$ and  $R_i^j$ oscillate in anti-phase but with a large period compared to the previous case with the bias.
The location of the resonances in $T_i^i$ (coincide with $R_i^j$=0 ) are given by
\begin{equation}
d_n=\frac{\pi n}{\sqrt{\left\vert \left(E-v_0\right)^2-\Omega^2 \right\vert}}.
\end{equation}
Note that these locations are energy dependent in contrary to those in the biased system shown in Fig. \ref{T_vs_d}(c). For example, in the energy interval  $0<E<v_0-\Omega$, as the energy increases,  fewer resonances appear in the  considered   interaction width. Most importantly, we observe that $T_i^i+T_i^j\neq1$ and $R_i^i+R_i^j\neq0$ are  always preserved unless $d=d_n$. In other words, the backscattering is not strictly prohibited as the case when Klein tunneling exists, but instead a none zero reflected current exists at specific widths.        Hence, we conclude that
 in the presence of the interlayer mass-term difference Klein tunneling is hampered, and instead Febry-P\'erot resonances appear. 

Next, we consider asymmetric mass-term on both layers such that $\Delta_0=0.3\gamma_1$ and $\Omega=0.4\gamma_1$ and show all possible transmission and reflection channels in Fig. \ref{T_R_Asymmetric}. The most remarkable feature is the asymmetric inter-cone transmission with respect to normal incidence such that $T_+^-(k_y)=T_-^+(-k_y)$. This asymmetry feature    is also strongly pronounced within the gap in all reflection channels. However, the inter-cone reflections are the same such that  $R_+^-(k_y)=R_-^+(k_y)$ since the carriers are reflected to the same region of incidence. This angular asymmetry manifests that the  Hamiltonian in Eq. \eqref{eq02} is not invariant under the  exchange $k_y\rightarrow-k_y$ and in the presence of $\Omega$ and $\Delta_0$. Note that breaking only the interlayer symmetry in AA-BLG by considering either $\delta\neq0$ or $\Omega\neq0$  leaves the  Hamiltonian in Eq. \eqref{eq02}  invariant under the  exchange $k_y\rightarrow-k_y$, and  thus results in symmetric transmission and reflection probabilities,  see  Figs. (\ref{T_Zero_Om}, \ref{T_R_Symmetric}). For example,  for $\delta\neq0$ we can show that $H(k_y)=UH(-k_y)$ where $U$ here is a unitary transformation corresponding to  an interchange of the $\phi_1$ and $\phi_2$ basis together with $\phi_3\leftrightarrow\phi_4$.   This is diametrically opposite to the AB-BLG\cite{Van_Duppen01_2013} system where breaking the interlayer symmetry destroys such invariance under  the exchange $k_y\rightarrow-k_y$. This discrepancy     stems from  the  symmetric  and asymmetric interlayer coupling in AA-BLG and AB-BLG  systems, respectively. Furthermore,
comparing Fig. \ref{T_R_Asymmetric} with the results in  Fig. \ref{T_R_Symmetric}, we see that here     channels carry out a non-zero current for normal incidence which is an extra consequence of breaking the angular symmetry. It is important to point out that in the vicinity of  the other valley, the inter-cone transmission is given by $T_i^j(k_y)_K=T_i^j(-k_y)_{K^{'}}$. This angular asymmetry; however, leaves the valley degeneracy unchanged and thus the overall symmetry of the system is preserved as well as the macroscopic time reversal symmetry. Eventually, it is worth to mention that the cloaking effect presents in gapless\cite{Gu2011} and gapped\cite{Van_Duppen01_2013} AB-BLG is absent in AA-BLG as can be deduced from Figs. (\ref{T_R_Symmetric}, \ref{T_R_Asymmetric}). 
\begin{figure}[tb]
\vspace{0.cm}
\centering\graphicspath{{./Figures/}}
\includegraphics[width=\linewidth]{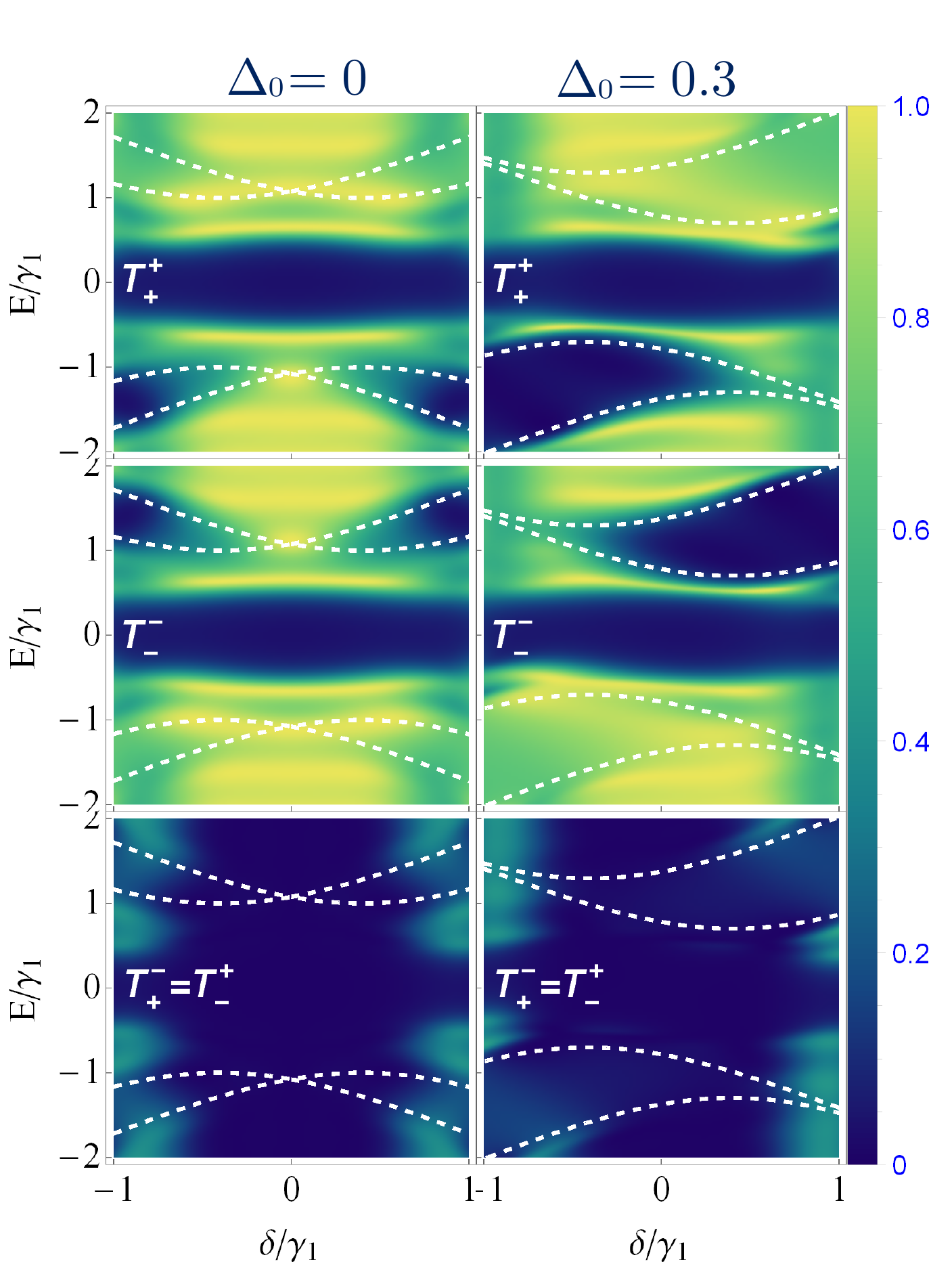}
\vspace{0.cm}
\caption{Density plot of  transmission probabilities  of  different channels as a function of the bias for $k_y=0,\ \Omega=0.4\gamma_1,\ v_0=0, \ d=6l.$  Positive and negative strengths of the bias correspond to electric field with opposite directions, the bands are superimposed  as white dashed curves.}\label{Electron_Hole_Symm}
\end{figure}
\begin{figure}[tb]
\vspace{0.cm}
\centering\graphicspath{{./Figures/}}
 \includegraphics[width=\linewidth]{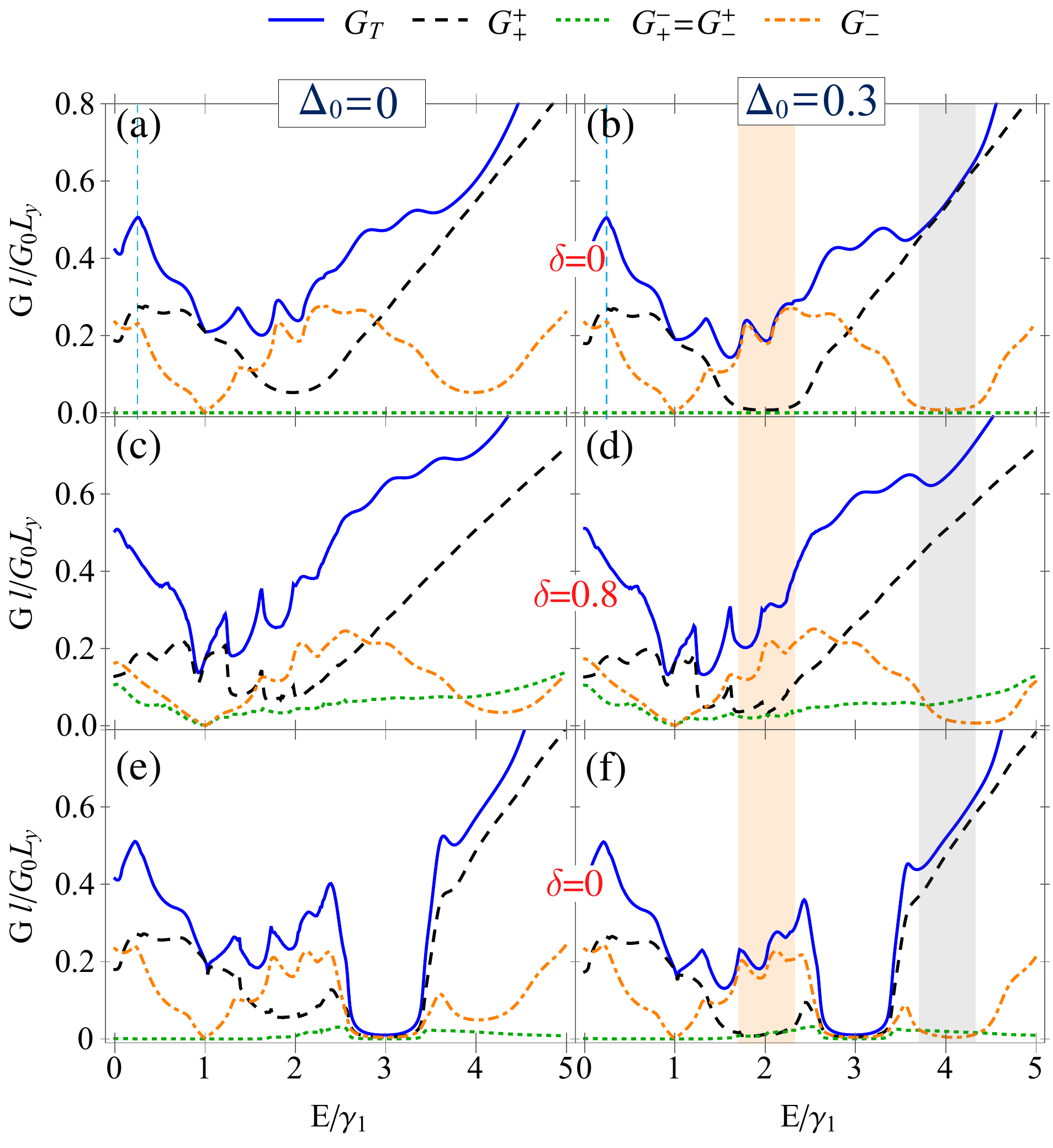}
\vspace{0.cm}
\caption{Contribution of  intra- and inter-cone conductances for $d=6l,\ v_0=3\gamma_1$ with  (a-d) $ \Omega=0$, and 
(e, f) $\Omega=0.4\gamma_1$. The pink and gray regions in the right panels represent the gap width in the vicinity of lower and upper Dirac cones , respectively, which induced by the mass-term amplitude $\Delta_0$. }\label{Conductance}
\end{figure}

In Fig. \ref{Electron_Hole_Symm}, we plot the variation of the inter- and intra-cone transmission probabilities, for normal incidence, in terms of the potential bias $\delta$ and Fermi Energy $E$   for zero and finite mass-term amplitude $\Delta_0$. When looking at the transmission profiles, one can distinguish two kinds of symmetry, namely, cone  and electron-hole symmetries. In the cone symmetry, the two intra-cones transmission  are connected through  $T_+^+(E,\delta)=T_-^-(- E,\tau\delta)$ with $\tau=(+,-)$ for $\Delta_0=0$ and $\Delta_0\neq0$, respectively. While the electron-hole symmetry is only preserved in the inter-cone transmission such that $T_+^-(E,-\delta)=T_+^-(- E,\delta)$.

 As a measurable quantity, we show in Fig. \ref{Conductance} the intra-, inter-cone and total conductances of the system. In all panels we consider an electrostatic potential of strength $v_0=3\gamma_1$ whose width is $d=6l$ and the rest of parameters  $\delta, \Delta_0,$ and $\Omega$ are varied. In all panels, we note that the total conductance is considrably large for $E=0$  in striking contrast to single layer graphene\cite{Pereira2010}  and AB-BLG\cite{Barbier2009}. This  originates from the availability of propagating states at $E=0$ corresponding to lower and upper cones in AA-BLG,  as attested by the plots of Fig. \ref{Bands}(a). In single layer graphene and AB-BLG, such propagating states are absent at $E=0$ and only through evanescent states a ballistic transport takes place\cite{Snyman_2007}. The overall conductance profiles in Figs. \ref{Conductance}(a, b) exhibit smoothed resonances, particulary for $E<v_0$,  inherited from the transmission resonances. For example, at low energy the intra-cone channel $T_-^-$ possesses resonances  of flat shape\cite{Abdullah2018a}  that appear as sharp peaks in the conductance. These peaks correspond to energy situated almost at the tail of the resonances in the channel $T_-^-$. Since the tail of the resonances coincide with  $\pm\pi/2$ angle of incidence, we can calculate the  peak    energies using Eq. \eqref{resonances} where it  can be rewritten for  $\delta=0$ as $E_n=1+ \sec^2\phi\left(  v_0-\sqrt{v_0^2\sin^2\phi+\kappa^2 \cos^2\phi} \right)$, with $\phi$ is the incident angle and $\kappa^2=[\Delta_0^2+(n\pi/d)^2]$. Now, we can see that for $\phi\rightarrow\pm\pi/2$, $E_7=(0.261,\ 0.246 )\gamma_1$ for $\Delta_0=(0, 0.3)\gamma_1$, respectively. These two peaks are superimposed as vertical dashed-blue lines in  Figs. \ref{Conductance}(a, b). Even though the total conductance profile $G_T$ remains almost unaffected with or without the mass-term amplitude, as can be inferred from Figs. \ref{Conductance}(a, b),  the intra-cone conductances $G_+^+$ and $G_-^-$  are drastically altered in the presence of mass term amplitude $\Delta_0$. In particular, they  are totally suppressed in the vicinity of the lower and upper cones as clarified by the  dashed-black and -blue curves in Fig. \ref{Conductance}(b) and stipulated by the energy spectrum in Fig. \ref{Bands}(b). Note that the inter-cone conductance $G_i^j$ is strictly zero as the two cones are being decoupled, see dashed-green curves in Figs. \ref{Conductance}(a, b).

 As shown in Figs. \ref{Conductance}(c, d), the potential bias  strongly modifies the positions
and shapes of the resonances in the conductance channels. Of particular importance, it switches on the inter-cone conductance ($G_i^j\neq0$), which in turn increases the total conductance and results in  pronounced resonances. 

In panels (e, f) of Fig. \ref{Conductance}, we show the conductance for finite  interlayer mass-term difference  with zero   potential bias.  The conductance profiles are almost similar to those in panels (a, b), respectively,  but are influenced by the suppression in the energy range of the band gap at $E=v_0\pm E_g/2$.  Since the inter-cone conductance $G_i^j$ is relatively small,     the resonances in the energy interval $v_0- E_g/2>E>\gamma_1$ are smeared. Furthermore,     we note that the total conductance is considrably large at the edges of the gap.  From an applied perspective,  this provides an efficient configuration   to \textit{switch on/off} the current through the sample using only an electrostatic gate.
The presence of the gap   also provides a characteristic signature of interlayer-mass term difference, which we expect to be elegantly observable in experiments.
According to the total conductance profiles in panels (e, f), we note that within the gap the total conductance is not strictly zero. This is because we consider a relatively narrow interaction region, $d=6l$, and thus evanescent waves can take part in the electronic transport\cite{Snyman_2007}. However , for $d\geqslant10l$ the total conductance within the gap  becomes completely zero.

\section{Summary and conclusion}\label{summary}

In conclusion,  we proposed a configuration to  establish a gap  in the energy spectrum of AA-BLG by  considering  dielectric-induced mass term.  We analytically derive the energy spectrum and  the wavefunction of the system with symmetric and asymmetric mass term and in the presence of an electrostatic bias. We have evaluated the quantum transport through the biased and gapped AA-BLG  system. Specifically, we scrutinized  chiral tunnelling properties of the charge carriers
in the presence of the  potential bias and  interlayer mass-term difference.
We found  that Klein tunneling was maintained with the bias but  the mass-term difference  completely destroyed it and instead  Febry-P\'erot resonances were surfaced. Furthermore,   we showed that both parameters breaks the interlayer symmetry and couple the two Dirac cones. This coupling established inter-cone scattering that is asymmetric with respect to normal incidence when considering asymmetric mass term on both layers. In gapped AA-BLG, the electron-hole symmetry was broken and new symmetries emerged with the intra- and inter-cone channels, namely, $T_+^+(E,\delta)=T_-^-(- E,\tau\delta)$  and $T_+^-(E,-\delta)=T_+^-(- E,\tau\delta)$ with $\tau=(+,-)$ for $\Delta_0=0$ and $\Delta_0\neq0$.  

For gated AA-BLG ($v_0\neq0$),  we showed that the mass term amplitude $\Delta_0$ slightly alters the  total conductance while drastically changes  the intra-cone conductances where they drop to zero in  the vicinity of the upper and lower cones.  On the other hand, the resulting conductance significantly increases in biased AA-BLG as a result of the extra inter-cone channels that can be accessed by the  bias. Consequently, the peaks in the total conductance become very pronounced and their locations are modified. Introducing the mass-term difference forms a distinct characteristic in the conductance represented by a   gap whose location can be modulated by the electrostatic gate $v_0$.
Finally, we expect that  the existence of  topological states within the gap in this system is of  great potential when considering a kink mass-term profile since they were  already observed in single layer graphene\cite{Zarenia_2012,Costa2017} whose spectrum resembles the one of  AA-BLG.
The results presented here are potentially
exploitable for paving  the way for    electrical control of quantum transport in  AA-BLG-based electronic devices. 
\section*{Acknowledgment}
We acknowledge the support of King Fahd University of Petroleum
and Minerals under research group project RG171007-1 \& RG171007-2. We
also acknowledge the material support of the Saudi Center for Theoretical
Physics (SCTP).  

\appendix
\section*{Appendices}
\section{Energy spectrum}\label{Appe-A}

The  energy spectrum  of the system can be obtained from Eq. \eqref{eq09} which gives 

\begin{eqnarray}
\epsilon_\alpha^{\pm}=\frac{1}{2}\left[\alpha\sigma
\pm \sqrt{-(\sigma^2+2\rho)- \textrm{ sgn}(\alpha)\frac{2\kappa}{\sigma}}\right]^{1/2},
\end{eqnarray}
where $\alpha=\pm1$ and
\begin{eqnarray*}
&&\sigma=\sqrt{\frac{1}{3}\left( -2\rho+\frac{C_0}{Q}+Q \right)},\\ 
&&Q=\frac{1}{2^{1/3}}\left(C_{1}  +\sqrt{C_{1}^2-4C_0^3}\right)^{1/3},\\ &&\rho=2\left[ \eta^2-2(\delta^2+1) \right],\ \kappa=8\Omega\delta\Delta_0,
\end{eqnarray*}
with
\begin{eqnarray*}
&&C_0=\rho^2+12\left[ \eta^{2}-4\Omega^2(\Delta^2-1) \right],\\ 
&&C_1=2\rho^3+27\kappa^2-72\rho\left[ \eta^2-4\Omega^2(\Delta^2-1) \right],\\ &&\eta^2=1+\delta^2-k_y^2-\Delta_0^2-\Omega^2. \end{eqnarray*}
\section{Wavefunction}\label{Appe-B}
The solution of Eq. \eqref{eq08} is a plane wave given by  
\begin{equation}\label{eqB1}
\phi_{2}=  e^{ik_{+}x}+e^{-ik_{+}x}+e^{ik_{-}x}+e^{-ik_{-}x}.
\end{equation}

Substituting this into equations (\ref{eq04}-\ref{eq07}) yields the  solutions.    The  wave function  of
the system can be written in  matrix form as
\begin{equation}\label{B2}
\Psi(x,y)=G M(x)C e^{ik_{y}y},
\end{equation}
where the four-component vector $C$ represents the different coefficients expressing the relative weights of the  different traveling modes, which have to be set according to the propagating region\cite{Van_Duppen01_2013}. The matrices $M(x)$ and $G$ are given by
\begin{equation}\label{B3}
M(x)=\textrm{Diag}[e^{ik_{+}x},e^{-ik_{+}x},e^{ik_{-}x},e^{-ik_{-}x}],
\end{equation}
and
\begin{equation}\label{B4}
G=\left(%
\begin{array}{cccc}
  \chi^{-}_{+} & \chi^{+}_{+} & \chi^{-}_{-} & \chi^{+}_{-} \\
  1 & 1 & 1 & 1 \\
  \zeta^{-}_{+} & \zeta^{+}_{+} & \zeta^{-}_{-} & \zeta^{+}_{-} \\
  \Lambda^{+} & \Lambda^{+} & \Lambda^{-} & \Lambda^{-}\\
\end{array}%
\right),
\end{equation}
where
\begin{eqnarray*}
&&\Lambda^{\pm}=\frac{1-k_y^2-k_{\pm}^{2}+-2(\epsilon+\Omega)+(\delta+\epsilon)^2- \lambda^2}{2\sqrt{2}(\epsilon+\Omega)},\\ \\
&&
\chi^{\pm}_{\alpha}=\pm\frac{(k_{\alpha}\pm ik_y)\left[ \sqrt{2}(\epsilon-\Delta_0)+\Lambda^{\alpha}(\mu+1) \right]}{1+(\delta+\Omega)^2-(\Delta_0-\epsilon)^2},\\ \\
 &&\zeta_{\alpha}^{\pm}=\pm\frac{(k_{\alpha}\pm ik_y)\left[ \sqrt{2}(\Omega+\delta)+\Lambda^{\alpha}(\mu-1) \right]}{1+(\delta+\Omega)^2-(\Delta_0-\epsilon)^2},
\end{eqnarray*}
with $\alpha=\pm$, $\lambda=\Delta_0-\Omega$, and $\mu=\delta-\lambda-\epsilon.$ Note that the above solution corresponds to the intermediate region $II$ shown in Fig. \ref{Device}(a). To obtain the desired solutions in regions $I$ and $III$ we need just to set $\delta=v_{0}=\Delta_0=\Omega=0$ in the above equations. Then, implementing the transfer matrix together with appropriate boundary conditions gives the transmission and reflection probabilities\cite{Barbier01_2010,Van_Duppen01_2013,Abdullah_2017}.



\end{document}